\documentclass[epj]{webofc}
\usepackage[varg]{txfonts}   
\wocname{EPJ Web of Conferences}
\woctitle{CONF12}
\def\Slash#1{\not\!\!#1}

\begin{document}
\selectlanguage{english}
\title{Some Relations for Quark Confinement and 
Chiral Symmetry Breaking in QCD}
%
%

\author{Hideo Suganuma\inst{1}\fnsep\thanks{\email{suganuma@scphys.kyoto-u.ac.jp}} \and
        Takahiro M. Doi \inst{1} 
\and
        Krzysztof Redlich \inst{2}
\and 
        Chihiro Sasaki \inst{2}
}

\institute{
Department of Physics, 
Kyoto University, Kitashirakawaoiwake, Sakyo, Kyoto 606-8502, Japan
\and
Institute of Theoretical Physics, University of Wroclaw, 
PL-50204 Wroclaw, Poland
}

\abstract{%
We analytically study the relation between quark confinement 
and spontaneous chiral-symmetry breaking in QCD.
In terms of the Dirac eigenmodes, we derive some formulae for 
the Polyakov loop, its fluctuations, and the string tension from the Wilson loop. 
We also investigate the Polyakov loop in terms of the eigenmodes of 
the Wilson, the clover and the domain wall fermion kernels, respectively.
For the confinement quantities, the low-lying Dirac/fermion eigenmodes are found to give 
negligible contribution, while they are essential for chiral symmetry breaking. 
These relations indicate no direct one-to-one correspondence 
between confinement and chiral symmetry breaking in QCD, 
which seems to be natural because confinement is realized 
independently of the quark mass. 
}
\maketitle
\def\slash#1{\not\!#1}
\def\slashb#1{\not\!\!#1}
\def\slashbb#1{\not\!\!\!#1}

\section{Introduction}

Color confinement and spontaneous chiral-symmetry breaking \cite{NJL61} 
are two outstanding nonperturbative phenomena in QCD, 
and they and their relation \cite{DSI14,DRSS15,SDI16} have been studied 
as one of the important difficult problems in theoretical particle physics.
For quark confinement, the Polyakov loop $\langle L_P \rangle$ is 
one of the typical order parameters, 
which relates to the single-quark free energy $E_q$ as 
$\langle L_P \rangle \propto e^{-E_q/T}$ at temperature $T$. 
For chiral symmetry breaking, the standard order parameter 
is the chiral condensate $\langle \bar qq \rangle$, 
and low-lying Dirac modes are known to play the essential role \cite{BC80}.

\begin{figure}[h]
\begin{center}
\includegraphics[scale=0.27]{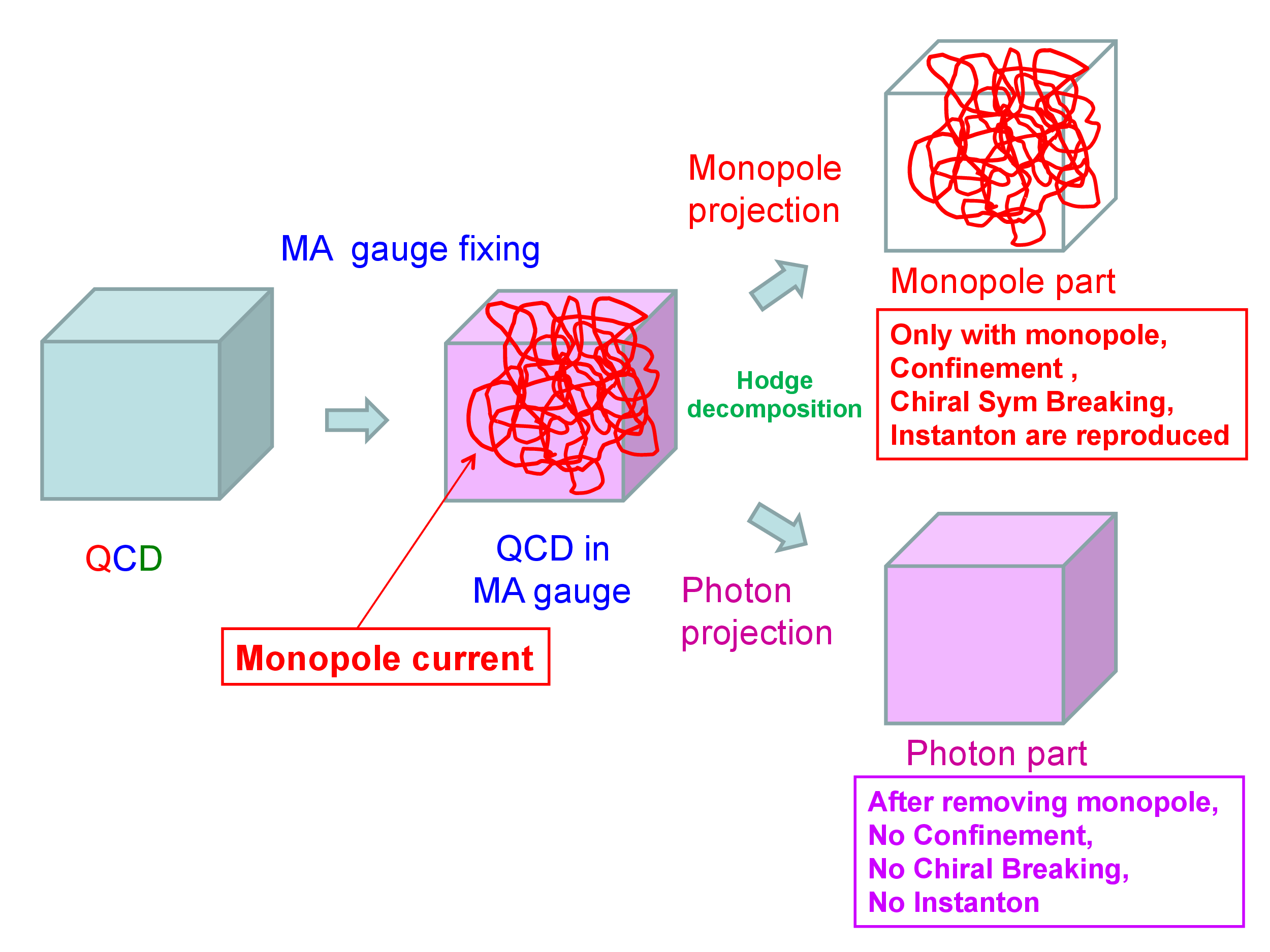}
\hspace{0.5cm} \includegraphics[scale=0.24]{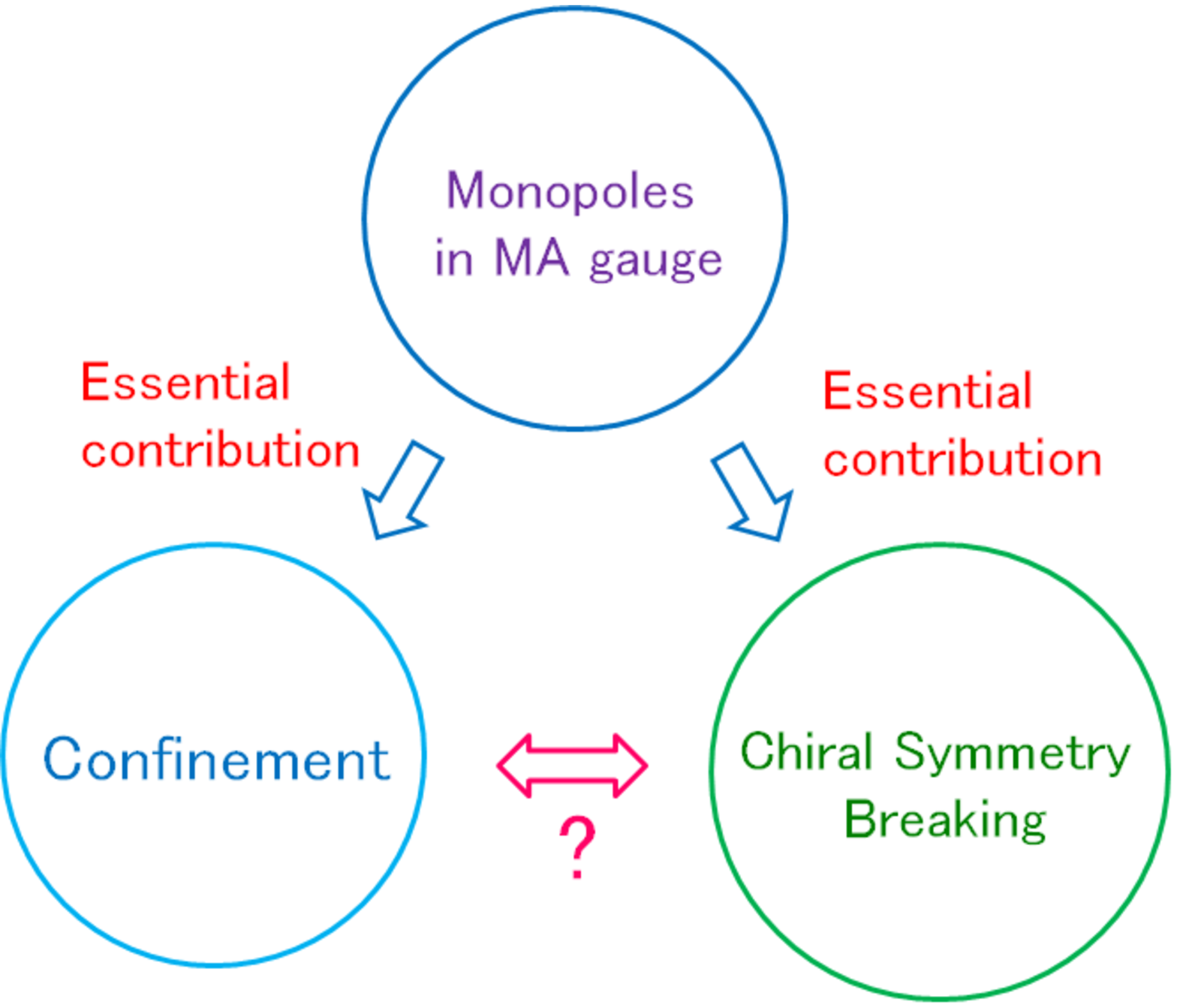}
\caption{
In the MA gauge, monopoles topologically appear. 
By removing the monopole from the QCD vacuum, 
confinement and chiral symmetry breaking are simultaneously lost \cite{SST95,M95W95}. 
This means crucial role of monopoles to both confinement and chiral symmetry breaking, but does not mean the direct correspondence between them.
}
\label{fig:MAG}
\end{center}
\vspace{-0.3cm}
\end{figure}

A strong correlation between confinement and chiral symmetry breaking 
has been suggested by approximate coincidence between deconfinement and 
chiral-restoration temperatures \cite{R12}.
Their correlation has been also suggested 
in terms of QCD-monopoles \cite{SST95,M95W95}, 
which topologically appear in QCD in the maximally Abelian (MA) gauge. 
By removing the monopoles from the QCD vacuum, 
confinement and chiral symmetry breaking 
are simultaneously lost in lattice QCD 
\cite{SST95,M95W95}. (See Fig.~\ref{fig:MAG}.)
This indicates an important role of QCD-monopoles 
to both confinement and chiral symmetry breaking, and thus 
these two phenomena seem to be related via the monopole.
However, the direct relation of confinement and chiral symmetry breaking 
is still unclear.

Actually, an accurate lattice QCD study \cite{AFKS06} shows 
about 25MeV difference between 
the deconfinement and the chiral-restoration temperatures, i.e., 
$T_{\rm deconf}\simeq 176 {\rm MeV}$ and $T_{\rm chiral}\simeq 151 {\rm MeV}$.
We also note that some QCD-like theories exhibit a large difference 
between confinement and chiral symmetry breaking. 
For example, in an SU(3) gauge theory with adjoint-color fermions, 
the chiral transition occurs at much higher temperature, 
$T_{\rm chiral} \simeq 8 T_{\rm deconf}$ \cite{KL99}.
In 1+1 QCD with $N_f \ge 2$, confinement is realized, 
but spontaneous chiral symmetry breaking does not occur, 
because of the Coleman-Mermin-Wagner theorem.
Also in $N=1$ SUSY 1+3 QCD with $N_f=N_c+1$,
while confinement is realized, chiral symmetry breaking does not occur.
A recent lattice study of SU(2)-color QCD with $N_f=2$ shows that 
a confined but chiral-restored phase is realized at a large baryon density \cite{BIKMN16}. 

In this paper, considering the essential role of 
low-lying Dirac modes to chiral symmetry breaking \cite{R12}, 
we derive analytical relations between the Dirac modes 
and the confinement quantities, e.g., the Polyakov loop \cite{DSI14}, 
its fluctuations \cite{DRSS15} and the string tension \cite{SDI16}, 
in the lattice QCD formalism.

\section{Dirac operator, Dirac eigenvalues and Dirac modes in lattice QCD}

In this paper, we take an ordinary square lattice with spacing $a$ and 
the size $V \equiv N_s^3 \times N_t$, and impose 
the standard temporal periodicity/anti-periodicity for gluons/quarks.
In lattice QCD, the gauge variable is expressed as 
the link-variable $U_\mu(s) \equiv {\rm e}^{iagA_\mu(s)}$
with the gauge coupling $g$ and the gluon field $A_\mu(x)$, and 
the simple Dirac operator and the covariant derivative operator are given as 
\begin{eqnarray}
\Slash{\hat D}
=\frac{1}{2a}\sum_{\mu=1}^{4} \gamma_\mu (\hat U_\mu-\hat U_{-\mu}), \quad
\hat D_\mu
=\frac{1}{2a} (\hat U_\mu-\hat U_{-\mu}),
\label{eq:Dirac}
\end{eqnarray}
where the link-variable operator 
$\hat U_{\pm \mu}$ \cite{DSI14,DRSS15,SDI16,GIS12} is defined by 
\begin{eqnarray}
\langle s |\hat U_{\pm \mu}|s' \rangle 
=U_{\pm \mu}(s)\delta_{s\pm \hat \mu,s'}, 
\label{eq:LVO}
\end{eqnarray}
with $U_{-\mu}(s)\equiv U^\dagger_\mu(s-\hat \mu)$.
This simple Dirac operator $\hat{\Slash D}$ is anti-hermite and satisfies 
\begin{eqnarray}
\hat{\Slash{D}}_{s',s}^\dagger=-\hat{\Slash{D}}_{s,s'}.
\end{eqnarray}
We define the normalized Dirac eigenmode $|n \rangle$ 
and the Dirac eigenvalue $\lambda_n$, 
\begin{eqnarray}
\hat{\Slash{D}} |n\rangle =i\lambda_n |n \rangle \quad (\lambda_n \in {\bf R}), 
\qquad
\langle m|n\rangle=\delta_{mn}.
\end{eqnarray}
Because of anti-hermiticity of $\hat{\Slash D}$, 
the Dirac eigenmode $|n \rangle$ satisfies the complete-set relation, 
\begin{eqnarray}
\sum_n |n \rangle \langle n|=1.
\label{eq:complete}
\end{eqnarray}

In lattice QCD, any functional trace becomes 
a sum over all the space-time site, i.e., ${\rm Tr} = \sum_s {\rm tr}$, 
which is defined for each lattice gauge configuration. 
For enough large volume lattice, e.g., $N_s \rightarrow \infty$, 
the functional trace is proportional to the gauge ensemble average, 
${\rm Tr}~\hat O =\sum_s {\rm tr}~\hat O
 \propto \langle \hat O \rangle_{\rm gauge~ave.}$, 
for any operator $\hat O$.
Note also that, because of the definition of $\hat U_{\pm \mu}$ in Eq.(\ref{eq:LVO}), 
the functional trace 
${\rm Tr}(\hat U_{\mu_1} \hat U_{\mu_2} \cdots \hat U_{\mu_N})$ 
of any product of link-variable operators 
corresponding to ``non-closed line'' is exactly zero 
\cite{DSI14,DRSS15,SDI16} 
at each lattice gauge configuration, 
before taking the gauge ensemble average.

In this paper, we mainly use the lattice unit, $a=1$, for the simple notation.

\section{Polyakov loop and Dirac modes in temporally odd-number lattice QCD}

To begin with, we study the Polyakov loop and Dirac modes 
in temporally odd-number lattice QCD \cite{DSI14,DRSS15,SDI16}, 
where the temporal lattice size $N_t(< N_s)$ is odd. 
In general, only gauge-invariant quantities 
such as closed loops and the Polyakov loop 
survive in QCD, according to the Elitzur theorem \cite{R12}.
All the non-closed lines are gauge-variant 
and their expectation values are zero.

Now, we consider the functional trace 
\cite{DSI14,DRSS15,SDI16}, 
\begin{eqnarray}
I \equiv {\rm Tr}_{c,\gamma}(\hat U_4\hat{\Slash{D}}^{N_t-1})
=\sum_n\langle n|\hat{U}_4\Slash{\hat{D}}^{N_t-1}|n\rangle
=i^{N_t-1}\sum_n\lambda_n^{N_t-1}\langle n|\hat{U}_4| n \rangle, 
\label{eq:FTI}
\end{eqnarray}
where ${\rm Tr}_{c,\gamma}\equiv \sum_s {\rm tr}_c 
{\rm tr}_\gamma$ 
includes the sum over all the four-dimensional site $s$
and the traces over color and spinor indices. 
In Eq.(\ref{eq:FTI}), we have used the completeness of the Dirac mode, 
$\sum_n |n\rangle \langle n|=1$.

\begin{figure}[h]
\begin{center}
\includegraphics[scale=0.45]{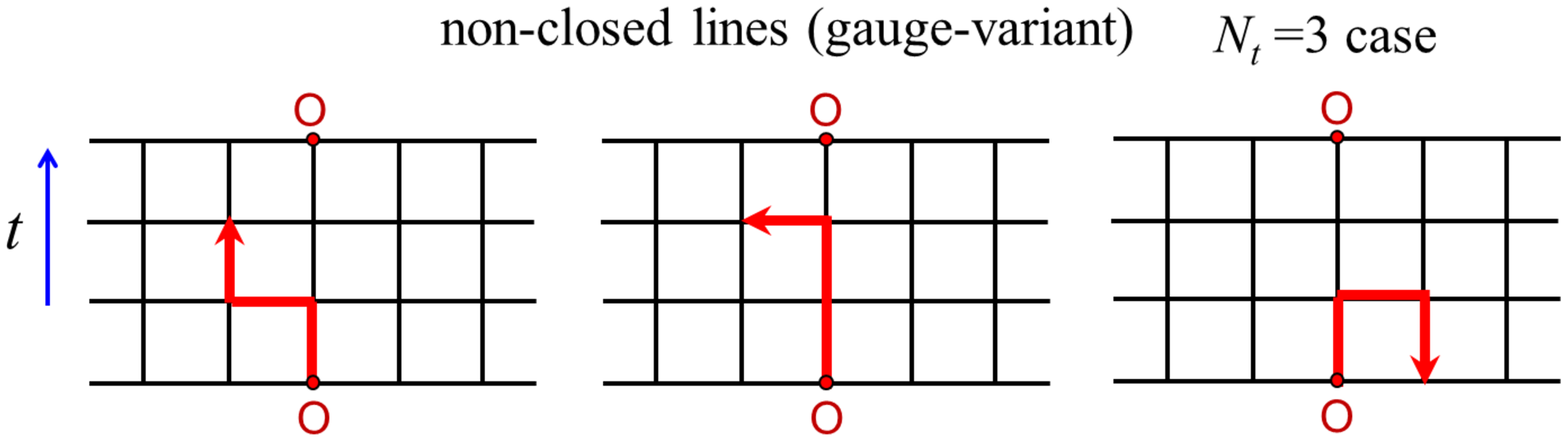}
\includegraphics[scale=0.285]{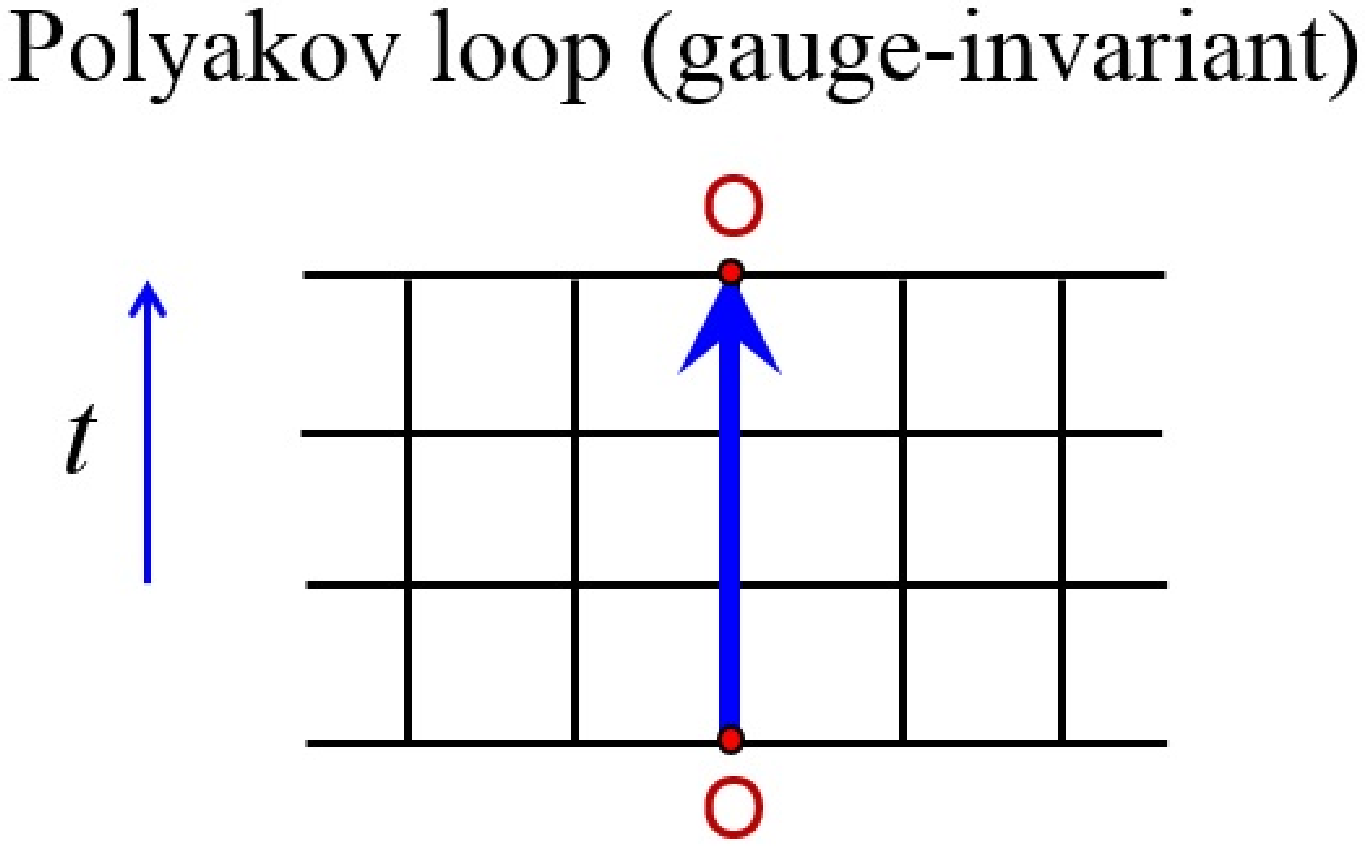}
\caption{
Partial examples of the trajectories in 
$I ={\rm Tr}_{c,\gamma}(\hat U_4\hat{\Slash{D}}^{N_t-1})$. 
For each trajectory, the total length is 
$N_t$, and the ``first step'' is positive 
temporal direction, $\hat U_4$.
All the trajectories with the odd length $N_t$ 
cannot form a closed loop on the square lattice,
so that they are gauge-variant and give no contribution, 
except for the Polyakov loop.
}
\label{fig:line}
\end{center}
\vspace{-0.3cm}
\end{figure}

From Eq.(\ref{eq:Dirac}), 
$\hat U_4\hat{\Slash{D}}^{N_t-1}$ 
is expressed as a sum of products of $N_t$ link-variable operators.
Then, $\hat U_4\hat{\Slash{D}}^{N_t-1}$ 
includes many trajectories with the total length $N_t$, 
as shown in Fig.~\ref{fig:line}.

Note that all the trajectories with the odd-number length $N_t$ 
cannot form a closed loop 
on the square lattice, and give gauge-variant contribution, 
except for the Polyakov loop.
Thus, in 
$I = {\rm Tr}_{c,\gamma}(\hat U_4\hat{\Slash{D}}^{N_t-1})$,
only the Polyakov-loop can survive 
as the gauge-invariant component, and 
$I$ is proportional to the Polyakov loop $L_P$.
Actually, we can mathematically derive 
the relation of 
\begin{eqnarray}
I = {\rm Tr}_{c,\gamma} (\hat U_4 \hat{\Slash D}^{N_t-1}) 
= {\rm Tr}_{c,\gamma} \{\hat U_4 (\gamma_4 \hat D_4)^{N_t-1}\} 
=4 {\rm Tr}_{c} (\hat U_4 \hat D_4^{N_t-1}) 
=
\frac{4}{2^{N_t-1}} {\rm Tr}_{c} \{ \hat U_4^{N_t} \} 
=-\frac{4N_cV}{2^{N_t-1}} L_P,
\end{eqnarray}
where the last minus reflects the temporal anti-periodicity of 
$\hat{\Slash{D}}$ \cite{SDI16}.
(Because of Eq.(\ref{eq:LVO}), 
the functional trace of ``non-closed line'' is exactly zero \cite{DSI14,DRSS15,SDI16} 
at each lattice gauge configuration.)

In this way, we obtain the analytical relation between the Polyakov loop $L_P$ 
and the Dirac modes in QCD on the temporally odd-number lattice 
\cite{DSI14,DRSS15,SDI16},
\begin{eqnarray}
L_P =-\frac{(2i)^{N_t-1}}{4N_cV}
\sum_n\lambda_n^{N_t-1}\langle n|\hat{U}_4| n \rangle,
\qquad
\langle L_P \rangle=-\frac{(2i)^{N_t-1}}{4N_cV}
\left\langle \sum_n\lambda_n^{N_t-1}\langle n|\hat{U}_4| n \rangle \right\rangle_{\rm gauge~ave.},
\label{eq:main}
\end{eqnarray}
which are mathematically robust. 
From Eq.(\ref{eq:main}), one can investigate each Dirac-mode contribution 
to the Polyakov loop.
As a remarkable fact, 
low-lying Dirac modes give negligible contribution to the Polyakov loop,
because of the suppression factor $\lambda_n^{N_t -1}$ in Eq.(\ref{eq:main}) 
 \cite{DSI14,DRSS15,SDI16}.
In lattice QCD calculations, we have numerically confirmed 
the relation (\ref{eq:main}) and tiny contribution of 
low-lying Dirac modes to the Polyakov loop 
in both confined and deconfined phases \cite{DSI14}. 

\section{Polyakov-loop fluctuations and Dirac eigenmodes}

\begin{figure}[h]
\begin{center}
\includegraphics[scale=0.9]{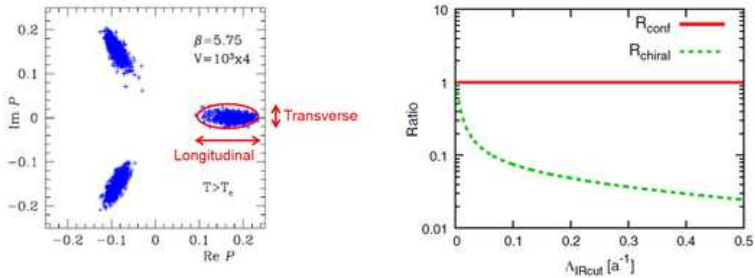}
\vspace{-0.3cm}
\caption{
(a) The scatter plot of the Polyakov loop in lattice QCD.
(b) The lattice QCD result for the infrared Dirac-mode cut quantities of 
$R_{\rm conf}(\Lambda_{\rm IRcut}) \equiv R_A(\Lambda_{\rm IRcut})/R_A$ 
and $R_{\rm chiral}(\Lambda_{\rm IRcut}) 
\equiv \langle \bar qq \rangle_{\Lambda_{\rm IRcut}}/\langle \bar qq \rangle$ 
plotted against the infrared cutoff $\Lambda_{\rm IRcut}$ introduced on 
Dirac eigenvalues \cite{DRSS15}. This figure is taken from Ref.\cite{DRSS15}.
In contrast to the sensitivity of the chiral condensate $R_{\rm chiral}$, 
the Polyakov-loop fluctuation ratio $R_{\rm conf}$ is almost unchanged 
against the infrared cutoff $\Lambda_{\rm IRcut}$ of the Dirac mode.
}
\label{fig:PF}
\end{center}
\vspace{-0.5cm}
\end{figure}

Next, we investigate Polyakov-loop fluctuations. 
As shown in Fig.~\ref{fig:PF}(a), the Polyakov loop $L_P$ has fluctuations 
in longitudinal and transverse directions.
We define its longitudinal and transverse components, 
$L_L \equiv {\rm Re}~\tilde L_P$ and $L_T \equiv {\rm Im}~\tilde L_P$,
with $\tilde L_P \equiv L_P~e^{2\pi i k/3}$ where $k \in \{0, \pm 1\}$ 
is chosen such that the $Z_3$-transformed Polyakov loop lies 
in its real sector \cite{DRSS15,LFKRS13}.
We introduce the Polyakov-loop fluctuations as 
$\chi_A \propto \langle |L_P|^2 \rangle-|\langle L_P \rangle|^2$, 
$\chi_L \propto \langle L_L^2 \rangle-\langle L_L \rangle^2$ and 
$\chi_T \propto \langle L_T^2 \rangle-\langle L_T \rangle^2$.
Some ratios of them largely change around the transition temperature, 
and can be good indicators of the QCD transition \cite{LFKRS13}. 

In temporally odd-number lattice QCD, 
we derive Dirac-mode expansion formulae for Polyakov-loop fluctuations \cite{DRSS15}.
For example, the Dirac spectral representation of the ratio 
$R_A \equiv \chi_A/\chi_L$ is 
\begin{align}
R_A=
\frac{
\left\langle\left|\sum_n\lambda_n^{N_t-1}\langle n|\hat{U}_4| n \rangle\right|^2\right\rangle_{\rm gauge~ave.}
-
\left\langle\left|\sum_n\lambda_n^{N_t-1}\langle n|\hat{U}_4| n \rangle\right|\right\rangle^2_{\rm gauge~ave.}
}{
\left\langle\left(\sum_n\lambda_n^{N_t-1}{\rm Re}\left(\mathrm{e}^{2\pi i k/3}\langle n|\hat{U}_4| n \rangle\right)\right)^2\right\rangle_{\rm gauge~ave.}
-
\left\langle\sum_n\lambda_n^{N_t-1}{\rm Re}\left(\mathrm{e}^{2\pi i k/3}\langle n|\hat{U}_4| n \rangle\right)\right\rangle^2_{\rm gauge~ave.}
}.
\end{align}
Because of the reduction factor $\lambda_n^{N_t-1}$ in the Dirac-mode sum, 
all the Polyakov-loop fluctuations are almost unchanged 
by removing low-lying Dirac modes \cite{DRSS15}.

As the demonstration, we show in Fig.~\ref{fig:PF}(b) the lattice QCD result 
of $R_{\rm conf}(\Lambda_{\rm IRcut}) \equiv R_A(\Lambda_{\rm IRcut})/R_A$ 
and $R_{\rm chiral}(\Lambda_{\rm IRcut}) 
\equiv \langle \bar qq \rangle_{\Lambda_{\rm IRcut}}/\langle \bar qq \rangle$ 
in the presence of the infrared Dirac-mode cutoff $\Lambda_{\rm IRcut}$ introduced on 
Dirac eigenvalues \cite{DRSS15}.
As the Dirac-mode cutoff $\Lambda_{\rm IRcut}$ increases, 
the chiral condensate $R_{\rm chiral}$ rapidly reduces, but 
the Polyakov-loop fluctuation ratio $R_{\rm conf}$ is almost unchanged \cite{DRSS15}.

\section{The Wilson loop and Dirac modes on arbitrary square lattices}

\begin{figure}[h]
\begin{center}
\includegraphics[scale=0.4]{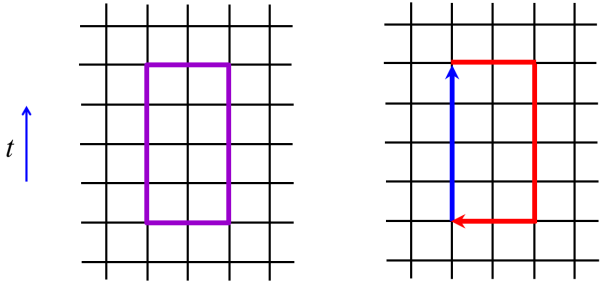}
\caption{Left: The Wilson loop $W$ on a $R \times T$ rectangle. 
Right: The factorization of the Wilson-loop operator 
as a product of 
$\hat U_{\rm staple}\equiv \hat U_i^R \hat U_{-4}^T \hat U_{-i}^R$ 
and $\hat U_4^T$ \cite{SDI16}. Here, $T$, $R$ and the lattice size 
$N_s^3 \times N_t$ are arbitrary.
}
\label{fig:WL}
\end{center}
\vspace{-0.3cm}
\end{figure}

In this section, we investigate the Wilson loop and the string tension 
in terms of the Dirac modes, on {\it arbitrary} square lattices 
with any number of $N_t$ \cite{SDI16}.
We consider the ordinary Wilson loop of the $R \times T$ rectangle.
The Wilson loop on the $x_i$-$t$ ($i=1,2,3$) plane is expressed by the functional trace, 
\begin{eqnarray}
W \equiv {\rm Tr}_{c} \hat U_i^R \hat U_{-4}^T \hat U_{-i}^R \hat U_4^T
={\rm Tr}_{c} \hat U_{\rm staple} \hat U_4^T,
\end{eqnarray}
where we introduce the ``staple operator'' $\hat U_{\rm staple}$ defined by 
\begin{eqnarray}
\hat U_{\rm staple} \equiv \hat U_i^R \hat U_{-4}^T \hat U_{-i}^R.
\end{eqnarray}
In fact, the Wilson-loop operator is factorized as 
a product of $\hat U_{\rm staple}$ and $\hat U_4^T$, 
as shown in Fig.~\ref{fig:WL}.

\subsection{Case of even T}

In the case of even number $T$, we consider the functional trace,
\begin{eqnarray}
J \equiv {\rm Tr}_{c,\gamma} \hat U_{\rm staple} \hat {\Slash D}^T
=\sum_{n} \langle n| \hat U_{\rm staple} {\Slash D}^T |n \rangle 
= (-)^{\frac{T}{2}}
\sum_{n} \lambda_n^T \langle n| \hat U_{\rm staple} |n \rangle,
\end{eqnarray}
where the completeness of the Dirac mode, 
$\sum_n |n\rangle \langle n|=1$, is used.
Similarly in Sec.~3, one finds 
\begin{eqnarray}
J =\frac{1}{2^T} {\rm Tr}_{c,\gamma} \hat U_{\rm staple} 
     \left[\sum_{\mu=1}^{4} \gamma_\mu (\hat U_\mu-\hat U_{-\mu})\right]^T
\!\!\! = \frac{1}{2^T} {\rm Tr}_{c,\gamma} \hat U_{\rm staple} (\gamma_4 \hat U_4)^T
 = \frac{4}{2^T} {\rm Tr}_{c} \hat U_{\rm staple} \hat U_4^T 
 = \frac{4}{2^T} W,
\end{eqnarray}
at each lattice gauge configuration. 
In fact, $\hat U_4$ must be selected 
in all the $\hat{\Slash D} \propto \sum_\mu \gamma_\mu (\hat U_\mu-\hat U_{-\mu})$ 
in $\hat{\Slash D}^T$, to form a loop in the functional trace. 
All other terms correspond to non-closed lines and give exactly zero, 
because of the definition of $\hat U_{\pm \mu}$ in Eq.(\ref{eq:LVO}).
Thus, we obtain \cite{SDI16}
\begin{eqnarray}
W = \frac {(-)^{\frac{T}{2}}2^T}{4}
\sum_{n} \lambda_n^T \langle n| \hat U_{\rm staple} |n \rangle.
\end{eqnarray}
Then, the inter-quark potential $V(R)$ is written as
\begin{eqnarray}
V(R) =-\lim_{T \to \infty} 
\frac{1}{T}{\rm ln} \langle W \rangle
= -\lim_{T \to \infty}\frac{1}{T}
{\rm ln} \left| \left\langle \sum_{n} 
(2 \lambda_n)^T \langle n| \hat U_{\rm staple} |n \rangle \right \rangle \right|, 
\end{eqnarray}
and the string tension $\sigma$ is expressed as 
\begin{eqnarray}
\sigma =-\lim_{R,T \to \infty} \frac{1}{RT}{\rm ln} \langle W \rangle
= -\lim_{R,T \to \infty}\frac{1}{RT}
{\rm ln} \left | \left \langle \sum_{n} 
\lambda_n^T \langle n| \hat U_{\rm staple} |n \rangle \right \rangle \right|.
\end{eqnarray}
Owing to the reduction factor $\lambda_n^T$ in the sum, 
the string tension $\sigma$, i.e., the confining force, 
is to be unchanged by the removal of 
the low-lying Dirac-mode contribution.

\subsection{Case of odd T}

In the case of odd number $T$, 
the similar results can be obtained by considering 
\begin{eqnarray}
J \equiv {\rm Tr}_{c,\gamma} 
\hat U_{\rm staple} \hat U_4 \hat {\Slash D}^{T-1}
=\sum_{n} \langle n| 
\hat U_{\rm staple} \hat U_4 {\Slash D}^{T-1} |n \rangle 
= (-)^{\frac{T-1}{2}}
\sum_{n} \lambda_n^{T-1} \langle n| \hat U_{\rm staple} \hat U_4 |n \rangle.
\end{eqnarray}
Actually, one finds 
\begin{eqnarray}
J = \frac{1}{2^{T-1}} {\rm Tr}_{c,\gamma} \hat U_{\rm staple} \hat U_4 
      \left[\sum_{\mu=1}^{4} \gamma_\mu (\hat U_\mu-\hat U_{-\mu})\right]^{T-1}
\!\!\! =\frac{1}{2^{T-1}}{\rm Tr}_{c,\gamma} \hat U_{\rm staple} 
     \hat U_4 (\gamma_4 \hat U_4)^{T-1} 
 = \frac{4}{2^{T-1}} W,
\end{eqnarray}
and obtains for odd $T$ the similar formula of \cite{SDI16}
\begin{eqnarray}
W = \frac {(-)^{\frac{T-1}{2}}2^{T-1}}{4}
\sum_{n} \lambda_n^{T-1} \langle n| \hat U_{\rm staple} \hat U_4|n \rangle.
\label{eq:Wilsonodd}
\end{eqnarray}
Then, the inter-quark potential $V(R)$ and the sting tension $\sigma$ 
are written as
\begin{eqnarray}
V(R) &=&-\lim_{T \to \infty} 
\frac{1}{T}{\rm ln} \langle W \rangle
= -\lim_{T \to \infty}\frac{1}{T}
{\rm ln} \left| \left \langle \sum_{n} 
(2 \lambda_n)^{T-1} \langle n| \hat U_{\rm staple} \hat U_4|n \rangle 
\right \rangle\right|, 
\cr
\sigma &=&-\lim_{R,T \to \infty} \frac{1}{RT}{\rm ln} \langle W \rangle
= -\lim_{R,T \to \infty}\frac{1}{RT}
{\rm ln} \left | \left \langle \sum_{n} 
\lambda_n^{T-1} \langle n| \hat U_{\rm staple} \hat U_4|n \rangle \right \rangle \right|,
\end{eqnarray}
where $\sigma$ is unchanged by removing 
the low-lying Dirac-mode contribution due to $\lambda_n^{T-1}$ 
in the sum.

\section{The Polyakov loop v.s. Wilson, clover and domain wall fermions}

All the above formulae are mathematically correct, 
because we have just used the Elitzur theorem 
(or precisely Eq.(\ref{eq:LVO}) for $\hat U_{\pm \mu}$) and 
the completeness $\sum_n|n \rangle \langle n|=1$ on the 
Dirac operator (\ref{eq:Dirac}).
However, one may wonder the doublers \cite{R12} in the use of 
the simple lattice Dirac operator (\ref{eq:Dirac}).
In this section, we express the Polyakov loop with the eigenmodes of 
the kernel of the Wilson fermion \cite{R12}, 
the clover ($O(a)$-improved Wilson) fermion \cite{SW85,NNMS03}
and the domain wall (DW) fermion \cite{K92,S93FS95}, respectively \cite{SDRS16}.
In these fermionic kernel, light doublers are absent.

\subsection{The Wilson fermion}

The Wilson fermion kernel \cite{R12} can be described with 
the link-variable operator $\hat U_{\pm \mu}$ as \cite{SDRS16}
\begin{eqnarray}
\hat K=\hat {\Slash D}+m+
\frac{r}{2a}\sum_{\mu=\pm 1}^{\pm 4} \gamma_\mu (\hat U_\mu-1)
=\frac{1}{2a}\sum_{\mu=1}^{4} \gamma_\mu (\hat U_\mu-\hat U_{-\mu})
+m+\frac{r}{2a}\sum_{\mu=1}^4 \gamma_\mu (\hat U_\mu+\hat U_{-\mu}-2),
\label{eq:WFK}
\end{eqnarray}
where each term of $\hat K$ includes one $\hat U_{\pm \mu}$ at most, 
and connects only the neighboring site or acts on the same site.
Near the continuum, $a \simeq 0$, 
Eq.(\ref{eq:WFK}) becomes $\hat K\simeq (\hat {\Slash D}+m)+ar \hat D^2$ 
and the Wilson term $ar \hat D^2$ is $O(a)$.

For the Wilson fermion kernel $\hat K$, 
we define its eigenmode $|n \rangle \rangle$ and 
eigenvalue $\tilde \lambda_n$ as 
\begin{eqnarray}
\hat K| n \rangle \rangle =i \tilde \lambda_n |n \rangle \rangle, \quad 
\tilde \lambda_n \in {\bf C}.
\end{eqnarray}
Note that, without the Wilson term, the eigenmode of $\hat K=\hat {\Slash D}+m$ is 
the simple Dirac eigenmode $|n\rangle$, i.e., $\hat K|n \rangle=(i\lambda_n+m)|n \rangle$, 
and satisfies the completeness of $\sum_n |n \rangle \langle n|=1$. 
In the presence of the $O(a)$ Wilson term, $\hat K$ is neither hermite nor anti-hermite, 
and the completeness may include an $O(a)$ error,
\begin{eqnarray}
\sum_n |n \rangle \rangle \langle \langle n|=1+O(a).
\label{eq:QCR}
\end{eqnarray}

\begin{figure}[h]
\begin{center}
\includegraphics[scale=0.5]{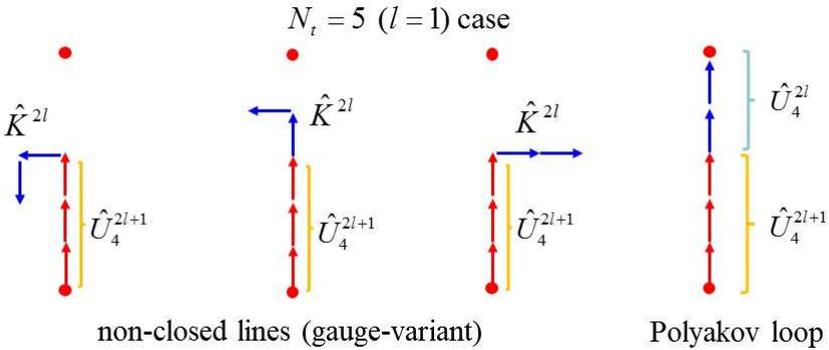}
\caption{
Some examples of the trajectories in 
$J \equiv {\rm Tr}(\hat U_4^{2l+1}\hat K^{2l})$ for the $N_t=5$ ($l=1$) case. 
The length does not exceed $N_t$ for each trajectory. 
Only the Polyakov loop $L_P$ can form a closed loop and survives in $J$. 
}
\label{fig:FK}
\end{center}
\end{figure}

Now, on the lattice with $N_t =4l+1$, we consider the functional trace, 
\begin{eqnarray}
J \equiv {\rm Tr}(\hat U_4^{2l+1} \hat K^{2l}).
\end{eqnarray}
Using the quasi-completeness of Eq.(\ref{eq:QCR}) for the eigenmode $|n \rangle\rangle$, 
one finds, apart from an $O(a)$ error, 
\begin{eqnarray}
J \simeq \sum_n\langle \langle n|\hat{U}_4^{2l+1}{\hat{K}}^{2l}|n \rangle \rangle
=\sum_n(i \tilde \lambda_n)^{2l} \langle \langle n|\hat{U}_4^{2l+1}| n \rangle \rangle.
\end{eqnarray}
We note that the kernel $\hat K$ in Eq.~(\ref{eq:WFK}) includes many terms, and 
$J \equiv {\rm Tr}(\hat U_4^{2l+1} \hat K^{2l})$ 
consists of products of link-variable operators, accompanying with $c$-number factors. 
In each product, 
the total number of $\hat U$ does not exceed $N_t$, because of Eq.~(\ref{eq:WFK}). 
Each product gives a trajectory as shown in Fig.~\ref{fig:FK}.

\noindent
Among the trajectories, however, 
only the Polyakov loop $L_P$ can form a closed loop 
and survives in $J$, so that one gets $J \propto L_P$. 
Thus, apart from an $O(a)$ error, we obtain \cite{SDRS16}
\begin{eqnarray}
L_P \propto
\sum_n \tilde \lambda_n^{2l}
\langle \langle n|\hat{U}_4^{2l+1}| n \rangle \rangle.
\end{eqnarray}
Due to the suppression factor of $\tilde \lambda_n^{2l}$ in the sum, 
one finds again small contribution from low-lying modes of $\hat K$ 
to the Polyakov loop $L_P$.

\subsection{The clover (O(a)-improved Wilson) fermion}

The clover fermion is an $O(a)$-improved Wilson fermion \cite{SW85}, and 
its kernel is expressed as \cite{SDRS16}
\begin{eqnarray}
\hat K=\frac{1}{2a}\sum_{\mu=1}^{4} \gamma_\mu (\hat U_\mu-\hat U_{-\mu})
+m+\frac{r}{2a}\sum_{\mu=1}^4 \gamma_\mu (\hat U_\mu+\hat U_{-\mu}-2)
+\frac{arg}{2}\sigma_{\mu\nu}G_{\mu\nu},
\label{eq:CFK}
\end{eqnarray}
with $\sigma_{\mu\nu}\equiv \frac{i}{2}[\gamma_\mu, \gamma_\nu]$. 
Here, $G_{\mu\nu}$ is the clover-type lattice field strength defined by 
\begin{eqnarray}
G_{\mu\nu}\equiv \frac{1}{8}(P_{\mu\nu}+P^\dagger_{\mu\nu}),
\end{eqnarray} 
with
\begin{eqnarray} 
P_{\mu\nu}(x) \equiv \langle x|
(\hat U_\mu \hat U_\nu \hat U_{-\mu} \hat U_{-\nu}
+ \hat U_\nu \hat U_{-\mu} \hat U_{-\nu} \hat U_\mu
+ \hat U_{-\mu} \hat U_{-\nu} \hat U_\mu \hat U_\nu
+ \hat U_{-\nu} \hat U_\mu \hat U_\nu \hat U_{-\mu} )
|x \rangle.
\end{eqnarray} 
The sum of the Wilson and the clover terms is $O(a^2)$ near the continuum, 
and the clover fermion gives accurate lattice results \cite{NNMS03}. 
Since $G_{\mu\nu}$ acts on the same site, 
each term of $\hat K$ in Eq.(\ref{eq:CFK}) 
connects only the neighboring site or acts on the same site, 
so that one can use almost the same technique as the Wilson fermion case. 
 
For the clover fermion kernel $\hat K$, 
we define its eigenmode $|n \rangle \rangle$ and 
eigenvalue $\tilde \lambda_n$ as 
\begin{eqnarray}
\hat K| n \rangle \rangle =i \tilde \lambda_n |n \rangle \rangle, \quad 
\tilde \lambda_n \in {\bf C}, \quad \qquad
\sum_n |n \rangle \rangle \langle \langle n|=1+O(a^2).
\label{eq:QCR2}
\end{eqnarray}
Again, on the lattice with $N_t =4l+1$, we consider the functional trace, 
\begin{eqnarray}
J \equiv {\rm Tr}(\hat U_4^{2l+1} \hat K^{2l}) 
\simeq \sum_n\langle \langle n|\hat{U}_4^{2l+1}{\hat{K}}^{2l}|n \rangle \rangle
=\sum_n(i \tilde \lambda_n)^{2l} \langle \langle n|\hat{U}_4^{2l+1}| n \rangle \rangle,
\end{eqnarray}
where we have used the quasi-completeness 
for $|n\rangle \rangle$ 
in Eq.(\ref{eq:QCR2}) within an $O(a^2)$ error. 
$J \equiv {\rm Tr}(\hat U_4^{2l+1} \hat K^{2l})$ 
consists of products of link-variable operators, 
accompanying with $c$-number factors, 
and each product gives a trajectory as shown in Fig.~\ref{fig:FK}.
Among the trajectories, 
only the Polyakov loop $L_P$ can form a closed loop 
and survives in $J$, i.e., $J \propto L_P$. 
Thus, apart from an $O(a^2)$ error, we obtain \cite{SDRS16}
\begin{eqnarray}
L_P \propto
\sum_n \tilde \lambda_n^{2l}
\langle \langle n|\hat{U}_4^{2l+1}| n \rangle \rangle, 
\end{eqnarray} 
and find small contribution from low-lying modes of $\hat K$ 
to the Polyakov loop $L_P$, due to $\tilde \lambda_n^{2l}$ in the sum.

\subsection{The domain wall (DW) fermion}

\begin{figure}[h]
\begin{center}
\includegraphics[scale=0.31]{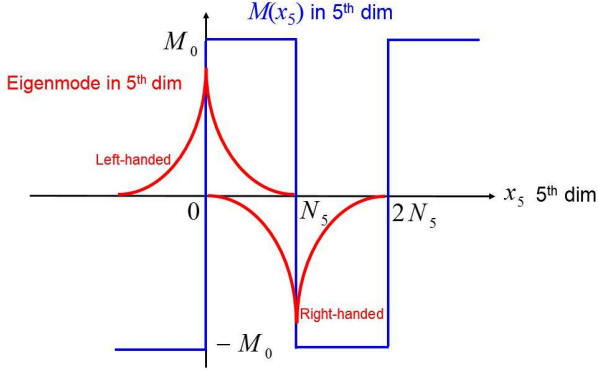}
\caption{
The construction of the domain wall (DW) fermion 
by introducing the fifth dimension of $x_5$ 
and the $x_5$-dependent mass $M(x_5)$. 
There appear left- and right-handed chiral zero modes localized 
around $x_5=0$ and $x_5=N_5$, respectively. 
}
\label{fig:DWF}
\end{center}
\vspace{-0.5cm}
\end{figure}

Finally, we consider the domain wall (DW) fermion \cite{K92,S93FS95}, 
where the ``exact'' chiral symmetry is realized in the lattice formalism 
by introducing an extra spatial coordinate $x_5$. 
The DW fermion is formulated in the five-dimensional space-time, 
and its (five-dimensional) kernel is expressed as 
\begin{eqnarray}
\hat K_5=\frac{1}{2a}\sum_{\mu=1}^{4} \gamma_\mu (\hat U_\mu-\hat U_{-\mu})
+m+\frac{r}{2a}\sum_{\mu=1}^4 \gamma_\mu (\hat U_\mu+\hat U_{-\mu}-2)
+\gamma_5\hat \partial_5+M(x_5),
\label{eq:DWFK}
\end{eqnarray}
where the last two terms in the RHS 
are the kinetic and the mass terms in the fifth dimension. 
Here, $x_5$-dependent mass $M(x_5)$ is introduced 
as shown in Fig.\ref{fig:DWF}, where $M_0=|M(x_5)|=O(a^{-1})$ is taken to be large. 
As for the extra coordinate $x_5$, there are only kinetic and mass terms in $\hat K_5$, 
so that the eigenvalue problem is solved in the fifth direction, 
and chiral zero modes are found to appear \cite{K92,S93FS95}.

For the five-dimensional DW fermion kernel $\hat K_5$, 
we define its eigenmode $|\nu \rangle$ and eigenvalue $\Lambda_\nu$ as 
\begin{eqnarray}
\hat K_5|\nu \rangle =i \Lambda_\nu |\nu \rangle, \quad 
\Lambda_\nu \in {\bf C}, \quad \qquad
\sum_\nu |\nu \rangle \langle \nu|=1+O(a).
\label{eq:QCR3}
\end{eqnarray}
Note that each term of $\hat K_5$ in Eq.(\ref{eq:DWFK}) 
connects only the neighboring site or acts on the same site 
in the five-dimensional space-time, 
and hence one can use almost the same technique as the Wilson fermion case. 
On the lattice with $N_t =4l+1$, we consider the functional trace, 
\begin{eqnarray}
J \equiv {\rm Tr}(\hat U_4^{2l+1} \hat{K}_5^{2l}) 
\simeq \sum_\nu \langle \nu|\hat{U}_4^{2l+1}{\hat{K}_5}^{2l}|\nu \rangle
=\sum_\nu(i \Lambda_\nu)^{2l} \langle \nu|\hat{U}_4^{2l+1}| \nu \rangle,
\end{eqnarray}
where the quasi-completeness for $|\nu \rangle$ in Eq.(\ref{eq:QCR3}) 
is used. 
$J \equiv {\rm Tr}(\hat U_4^{2l+1} \hat K_5^{2l})$ 
consists of products of link-variable operators with other factors, 
and each product gives a trajectory as shown in Fig.~\ref{fig:FK} 
in the projected four-dimensional space-time.
Among the trajectories, 
only the Polyakov loop $L_P$ can form a closed loop 
and survives in $J$, i.e., $J \propto L_P$. 
Thus, apart from an $O(a)$ error, we obtain 
\begin{eqnarray}
L_P \propto
\sum_\nu \Lambda_\nu^{2l} \langle \nu |\hat{U}_4^{2l+1}| \nu \rangle. 
\label{eq:DWPL}
\end{eqnarray} 

Because of the simple $x_5$-dependence in $\hat K_5$ in Eq.(\ref{eq:DWFK}), 
the extra degrees of freedom in the fifth dimension 
can be integrated out in the generating functional, and one obtains 
the four-dimensional physical-fermion kernel $\hat K_4$ \cite{S93FS95}. 
The physical fermion mode is given by 
the eigenmode $|n \rangle \rangle$ of $\hat K_4$, 
\begin{eqnarray}
\hat K_4 | n \rangle \rangle =i \tilde \lambda_n |n \rangle \rangle, \quad 
\tilde \lambda_n \in {\bf C}.
\end{eqnarray}
We find that the four-dimensional physical fermion eigenvalue 
$\tilde \lambda_n$ of $\hat K_4$
can be approximated by the eigenvalue 
$\Lambda_\nu$ of the five-dimensional DW kernel $\hat K_5$ as 
\begin{eqnarray}
\Lambda_\nu = \tilde \lambda_{n_\nu} +O(M_0^{-2})
=\tilde \lambda_{n_\nu} +O(a^2), 
\end{eqnarray}
where $M_0=|M(x_5)|=O(a^{-1})$ is taken to be large.

Combining with Eq.(\ref{eq:DWPL}), apart from an $O(a)$ error, we obtain 
\begin{eqnarray}
L_P \propto
\sum_\nu \tilde \lambda_{n_\nu}^{2l} 
\langle \langle \nu |\hat{U}_4^{2l+1}| \nu \rangle \rangle, 
\end{eqnarray} 
and find small contribution from low-lying physical-fermion modes of $\hat K_4$ 
to the Polyakov loop $L_P$, 
because of the suppression factor $\tilde \lambda_{n_\nu}^{2l}$ in the sum.

\section{Summary and Concluding Remarks}

In QCD, we have derived analytical relations between 
the Dirac modes and the confinement quantities such as 
the Polyakov loop, its fluctuations and the string tension in the lattice formalism, 
and have found negligible contribution from the low-lying Dirac modes to the confinement quantities. 

We have also investigated the Polyakov loop in terms of the eigenmodes of 
the Wilson, the clover and the domain wall fermion kernels, respectively, 
and have obtained the similar results.

These relations indicate no direct one-to-one correspondence 
between confinement and chiral symmetry breaking.
In other words, there is some independence 
of confinement from chiral properties in QCD.
This seems natural 
because confinement is realized independently of the quark mass. 

\section*{Acknowledgments}
H.S. would like to thank Prof. K. Konishi for the useful discussions, especially on SUSY QCD. 
H.S. and T.M.D. are supported by 
the Grants-in-Aid for Scientific Research (Grant No. 15K05076, 
No. 15J02108) from Japan Society for the Promotion of Science.
The work of K.R. and C.S. is partly
supported by the Polish Science Center (NCN) 
under Maestro Grant No. DEC-2013/10/A/ST2/0010.

\end{document}